\documentclass[preprint,12pt,showpacs]{revtex4} 
\usepackage{amsmath}
\usepackage{latexsym}
\usepackage{amssymb}
\usepackage{graphics,epstopdf}
\usepackage{amsmath,amssymb,amsthm}
\usepackage{slashed}
\usepackage[colorlinks=true, citecolor=blue, urlcolor=blue ]{hyperref}

\begin{document}
\title{Constraints on the choice of position-dependent effective mass and external potential for the existence of Lewis-Riesenfeld invariance and quantum canonical transformation }

\author{Kalpana Biswas}
\email{klpnbsws@gmail.com}
\affiliation{Department of Physics, University of Kalyani, West Bengal, India-741235}
\affiliation{Department of Physics, Sree Chaitanya College, Habra, North 24 Parganas, West Bengal-743268}
\author{Jyoti Prasad Saha}
\email{jyotiprasadsaha@gmail.com}
\affiliation{Department of Physics, University of Kalyani, West Bengal, India-741235}
\author{Pinaki Patra}
\thanks{Corresponding author}
\email{monk.ju@gmail.com}
\affiliation{Department of Physics, Brahmananda Keshab Chandra College, Kolkata, India-700108}

\date{\today}

\begin{abstract}
 Solving an arbitrary time-dependent system with position and time-dependent effective mass (TDPDEM) is an old challenge. Lewis-Riesenfeld -Ermakov's (LR) phase-space invariant method (LRIM) is an effective tool to handle time-dependent quantum systems. For Position-dependent effective-mass, the success of LRIM is limited.
   In this article, we have extended the  Lewis-Riesenfeld -Ermakov's phase-space invariant method for the general quantum system. We have obtained the restrictions on the choice of the position-dependent effective mass (PDEM), for which the LR-invariant operator will be of the close form. It turns out that, the choice of external potentials are also restricted for the existence of close form LR-invariant operator.  
   A class of unitary time-dependent quantum canonical transformation for the concerned PDEM and external potentials is presented, so that an equivalent time-independent PDEM Hamiltonian is obtained. 
   \end{abstract}
\maketitle
\section{Introduction}
 Lewis Riesenfeld (LR) invariant method \cite{LR1,LR2,LR3,LR4,LR5,LR6}  is an effective tool to construct the exact solutions, especially the coherent state (CS) structure of a time-dependent (TD) quantum system \cite{perturbation1,tdmagnetic1,quadratic hamiltonian,Landau1,perturbation2}. The key point of the LR invariant method (LRIM) is the existence of a time-invariant operator $ (\hat{\mathcal{I}}(t))$ on the phase space. Up to a time-dependent phase factor $e^{i\theta(t)}$, the eigenvectors $\left\{ \vert \psi_k (t) \rangle \right\}$ of TD phase-space invariant operator $ \hat{\mathcal{I}}(t)$ will satisfy the TD Schr\"{o}dinger equation corresponding to the Hamiltonian $\hat{H}(t)$.   However, the eigenvalues of $ \hat{\mathcal{I}}(t)$ are time-independent.  It is evident that the close form of $ \hat{\mathcal{I}}(t)$ in terms of the basic constituent operators such as position $(\hat{x})$, momentum $(\hat{p})$ and the combinations of powers of $\hat{x}$ and  $\hat{p}$ can be obtained for very specific class of Hamiltonians. Even for the classical systems, the Hamiltonians which are either linear or quadratic in momentum $(p)$ exhibit the LR invariants only for a specific class of the potential $V(q,t)$ functions \cite{LR4,LR5,perturbation2,Ermakov,Lewis1,Lewis2}.  In that sense, the studies of classical systems in this regard were not very much different than that was started by V. P. Ermakov for time-dependent harmonic oscillator \cite{Ermakov} which was subsequently generalized by Lewis for classical and quantum time-dependent harmonic oscillators \cite{Lewis1,Lewis2}. Classical conservation laws behind the classical LR-invariance were also well studied with the help of Noether's theorem \cite{Noether1,Noether2,Noether3}.\\
 However, most of the studies of the LR phase-space invariant method were confined to the constant mass system both in the classical and quantum mechanical scenario. There is no general study of LR-invariance for a quantum system with the position-dependent effective mass (PDEM) in the literature.  
 The concept of PDEM appears in diverse branches of Physics  \cite{yu,lozada,arias,guedes,mello,santos,cavalcanti,cunha,bekke,vitoria,vitoria1,bekke1,delta1}. For example, PDEM appears in the description of nonlinear optical properties in quantum well \cite{optical properties,optical properties2}, the asymmetric shape of crackling noise pulses emitted by a diverse range of noisy systems \cite{crackling noise,crackling2}, the cosmological models, even in quantum information theory \cite{qinfo1,qinfo2}. The appearance of PDEM in such diverse regime indicates that seemingly nonrelated phenomenon may be unified by the existence of PDEM.\\
Therefore, it is worth to revisit the concept of PDEM, especially under the shade of LR-invariance.
 The present article aims to provide the general class of position-dependent effective mass (PDEM) \cite{yu,lozada,arias,guedes,mello,santos,cavalcanti,cunha,bekke,vitoria,vitoria1,bekke1,delta1,optical properties,optical properties2,crackling noise,crackling2,qinfo1,qinfo2,CostaFilho,Muharimousavi,Souza Dutra,Souza Dutra1,Schmidt,Abdalla,Jha,heterostructure1,heterostructure2,Mario,pdem11,pinaki1} along with a general class of potentials for which close form LR-invariant operators can be constructed by the direct method. We have also shown that only a specific class of position-dependent mass profile and specific class of potentials can exhibit the close form of LR-invariance for the PDEM system. In particular, only a mass profile which varies with position $(x)$ as $\sim \frac{1}{x+a}$ under either Coulomb potential or potential well/barrier are allowed for the existence of a close form LR-invariant operator (LRIO). Interestingly, time dependent unitary quantum canonical transformation (TDQCT) \cite{qct1,qct2,qct3,qct4,qct5,qct6,qct7,qct8,qct9} exists for the PDEM systems for which LRIO is of the close form. The time dependent Hamiltonian for the concerned PDEM can be transformed into an equivalent time independent Hamiltonian under  TDQCT. \\
At first, we have demonstrated the conditions for which one can have a close form LR-invariant. Then the invariant operator  $ (\hat{\mathcal{I}}(t)) $ is constructed by LRM. Subsequently, a possible time-dependent canonical transformation is demonstrated, with the help of which one can prepare the system in an equivalent time-independent Hamiltonian.

\section{Position dependent effective mass Hamiltonian}
Due to the noncommutativity of position dependent effective mass (PDEM) and momentum, it was realized from the inception of the PDEM, that the kinetic part of a viable Hamiltonian for PDEM might be of the form (for simplicity, we are considering one-dimensional case) \cite{heterostructure1,heterostructure2}.
\begin{equation}\label{heterostructurehamiltonian}
\hat{H}= -\frac{1}{2} m^{\tilde{\alpha}} \left(\frac{d}{dx}\right)m^{\tilde{\beta}} \left(\frac{d}{dx}\right)m^{\tilde{\alpha}} .
\end{equation}
With the constraint on the constant parameters $\tilde{\alpha}$ and $\tilde{\beta}$
\begin{equation}\label{constraintalpha}
\tilde{\beta} + 2 \tilde{\alpha} =-1.
\end{equation}
Lack of consensus for the values of the parameters $\tilde{\alpha}$ and $\tilde{\beta}$ demands a careful consideration for the kinetic part of the Hamiltonian of PDEM. The consideration of the combinations of position and momentum from the classical level Legendre transformation between Lagrangian and Hamiltonian provides a viable form of kinetic part of hamiltonian as follows.
 $\{\vert x\rangle\}$ representation, the viable form of the Hamiltonian for arbitrary position and time dependent effective mass $(m(x,t))$ may be written as \cite{pinaki1}
\begin{equation}\label{quantum hamiltonian}
\hat{H}=-\frac{\partial}{\partial x}\left(\frac{1}{2m}\frac{\partial}{\partial x}\right) - \frac{1}{2m}\left(\frac{m'}{m}\right)^2 + V(x,t).
\end{equation} 
Prime denotes the derivative with respect to $x$. $V(x,t)$ is the potential function. Here we have considered $\hbar=1$ which will be followed throughout this article unless otherwise specified. \\
One can verify that under the usual boundary condition (wave functions vanish at the boundaries), the Hamiltonian \ref{quantum hamiltonian}  is self-adjoint for real-valued $m(x,t)$.
The occurrence of the intrinsic potential like function $-\frac{1}{2m}  \left(\frac{m'}{m}\right)^2$ appears solely due to the position dependency of the mass. In our discussion, the concept of mass is nothing but the measure of inertia. Hence, one can use the terminology "inertia-potential" $( V_{inertia})$ for the extra position and time-dependent function
\begin{equation}
V_{inertia} = -\frac{1}{2m}  \left(\frac{m'}{m}\right)^2 .
\end{equation} 
We can consider all the momentum independent terms altogether and define the effective potential $(v_{eff})$ as follows
\begin{equation}\label{veffective}
v_{eff}=   V_{inertia} + V(x,t) . 
\end{equation}
In the next section, we have demonstrated the restrictions on $m(x,t)$ and $V(x,t)$ such that the system posesses a LR-invariant operator.

\section{Restrictions on $m(x,t)$ and $V(x,t)$}
In this section, we have demonstrated the restrictions on PDEM $m(x,t)$ and the external potential $V(x,t)$ for which the system will have an LR-invariant operator. 
An operator  $ \hat{\mathcal{I}}(t)$ is invariant means
\begin{equation}\label{invariant condition}
\dot{\hat{\mathcal{I}}}(t) =\frac{\partial \hat{\mathcal{I}}(t)}{\partial t} + \frac{1}{i} \left[\hat{\mathcal{I}}(t), \hat{H}\right] =0 .
\end{equation}
Here dot $(^.)$ denotes the derivative with respect to time. We shall use this shorthand notation throughout this article unless otherwise specified.
Now the existence of an close form LR-invariant $\hat{\mathcal{I}}(t)$ based on the existence of finite number of generators $(\hat{\mathcal{O}}_i)$ of the quasi-algebra with respect to the Hamiltonian $\hat{H}$ such that the equation \ref{invariant condition} is satisfied. In particular we seek for an invariant operator of the form 
\begin{equation}
\hat{\mathcal{I}}(t) =\sum_{j=0}^N \mu_{j} (t) \hat{\mathcal{O}}_j , 
\end{equation}
such that the following quasi-algebra is closed for finite $N$.
\begin{equation}
\left[\hat{H}, \hat{\mathcal{O}}_i\right]= \sum_{k=1}^{N} \nu_{kj} \hat{\mathcal{O}}_j; \;\; i=1,...N.
\end{equation}
Where $\nu_{kj}$ are the structure constants of the algebra and  $\mu_{j}$ are arbitrary functions of time.\\
Our Hamiltonian is self-adjoint and quadratic in momentum. Therefore it is sufficient to consider the following operators as the generators of the algebra.
\begin{eqnarray}
\begin{array}{ccc}
\hat{\mathcal{O}}_1 = p ; &  \hat{\mathcal{O}}_2 = x ;  & \hat{\mathcal{O}}_3 =p \frac{1}{2m} ; \\
\hat{\mathcal{O}}_4 = \frac{1}{2m}p ; & \hat{\mathcal{O}}_5 = u(x) ; & \hat{\mathcal{O}}_6 = \frac{1}{2m} ; \\
\hat{\mathcal{O}}_7 = p \frac{1}{2m} p . & & 
\end{array}
\end{eqnarray}
Where u(x) represent any function of x.
$\hat{\mathcal{O}}_1$ and $\hat{\mathcal{O}}_2$ represents the simplest form namely linear dependency on momentum and co-ordinates respectively. $\hat{\mathcal{O}}_3$ and $\hat{\mathcal{O}}_4$ are chosen in such a way that the possible combinations of mass and momentum appears in equal footing. One can note that $\frac{1}{2} \left(\hat{ \mathcal{O}}_3 + \hat{\mathcal{O}}_4 \right)$ is self-adjoint. 
The commutation relations of these operators with the Hamiltonian are the followings.
\begin{eqnarray}\label{commutation with H}
\begin{array}{c}
\left[\hat{H}, \hat{\mathcal{O}}_1\right]=  -ip \frac{m'}{2m^2}p + i v_{eff}' ;\\
\left[\hat{H}, \hat{\mathcal{O}}_2\right]= -i\hat{\mathcal{O}}_3 - i\hat{\mathcal{O}}_4 ;\\
\left[\hat{H}, \hat{\mathcal{O}}_3\right]= -ip \left[\frac{m'}{2m^2} \frac{1}{2m}p,  \frac{1}{2m}p \frac{m'}{2m^2}\right] + \\ ip \frac{m'}{2m^2}\frac{1}{2m}p  + i v_{eff}' \frac{1}{2m} ; \\
\left[\hat{H}, \hat{\mathcal{O}}_4\right]= -i\frac{m'}{2m^2} \hat{\mathcal{O}}_7 + \frac{i}{2m}u' ;\\
\left[\hat{H}, \hat{\mathcal{O}}_5\right]= -i\hat{\mathcal{O}}_3 u' - i u' \hat{\mathcal{O}}_4 ;\\
\left[\hat{H}, \hat{\mathcal{O}}_6\right]=  i\hat{\mathcal{O}}_3 \frac{m'}{2m^2} + i \frac{m'}{2m^2}\hat{\mathcal{O}}_4 ; \\
\left[\hat{H}, \hat{\mathcal{O}}_7\right]=  i\hat{\mathcal{O}}_3 u' + i u' \hat{\mathcal{O}}_4. 
\end{array}
\end{eqnarray}
The structure constant of an algebra should be c-number. Therefore, the commutation of $\hat{\mathcal{O}}_1$ with $H$ suggests that $\hat{\mathcal{O}}_1$ must be discarded from the set of the algebra. Since, it makes the appearance of new operator (except for the case of constant mass which is trivial one) in the algebra. If one try to incorporate this new operator in the algebra, it is evident that this new operator will successively generate another independent operator. Similar situation occurs for the commutation  $\left[\hat{H}, \hat{\mathcal{O}}_3\right]$. Therefore the only choice for which the structure constants of the quasi-algebra ~\ref{commutation with H}
become c-number and the algebra contains finite number of basis is to remove $\hat{\mathcal{O}}_1$ from the set. Also, we have to restrict the possible forms of $u(x)$ which should be 
\begin{equation}
u(x)= v_{eff} . 
\end{equation}
Further, we have to put the constraint on the choice of $m(x)$ and $v_{eff}$ as follows.
\begin{eqnarray}\label{constraint on mv}
\frac{m'}{2m^2}= \alpha,\;\;\; v_{eff}'= \beta .
\end{eqnarray}
Where $\alpha, \beta$ are constants.\\
Therefore, in order to have the invariant $\hat{\mathcal{I}}$ in phase space for the position dependent effective mass (PDEM), only the following forms of PDEM and potential are allowed.
\begin{eqnarray}
m(x)= \frac{m_0}{1+l^{-1}x}. \label{choice of m}\\
V(x,t)= V_0 + \beta x + \frac{2\alpha^2 m_0}{1+l^{-1}x}. \label{choice of v}
\end{eqnarray}
Where $m_0= -\frac{1}{2c_0},\;\; l= \frac{c_0}{\alpha}$. $c_0$ and $V_0$ are the integration constants. \\
Therefore, we have seen that the only viable choice of mass and potentials are of the form \ref{choice of m} and \ref{choice of v} respectively. This is one of the important findings of the present study. In the following section, the explicit form of  $\hat{\mathcal{I}}(t)$ is demonstrated.
\section{Invariant operator}
From the preceding section, we have seen that $\mathcal{A}=\{\hat{H},\hat{\mathcal{O}}_2, \hat{\mathcal{O}}_3, \hat{\mathcal{O}}_4, \hat{\mathcal{O}}_6, \hat{\mathcal{O}}_7\}$ forms a quasi algebra. To envisage the proper forms of the structure constants, we have written the commutation relations explicitly as follows.
\begin{eqnarray}\label{structure constants}
\left[\hat{H},\hat{\mathcal{O}}_2\right] = -i\hat{\mathcal{O}}_3 - i\hat{\mathcal{O}}_4; \\
\left[\hat{H},\hat{\mathcal{O}}_3\right] = i\alpha\hat{\mathcal{O}}_7 +i\beta \hat{\mathcal{O}}_6 ; \\
\left[\hat{H},\hat{\mathcal{O}}_4\right] =- i\alpha\hat{\mathcal{O}}_7 +i\beta \hat{\mathcal{O}}_6 ; \\
\left[\hat{H},\hat{\mathcal{O}}_6\right] = i\alpha\hat{\mathcal{O}}_4 +i\alpha \hat{\mathcal{O}}_3 ; \\
\left[\hat{H},\hat{\mathcal{O}}_7\right] = i\beta\hat{\mathcal{O}}_3 +i\beta \hat{\mathcal{O}}_4 .
\end{eqnarray}
We can assume the form of $\hat{\mathcal{I}}$ as a linear combination of the generators of the algebra. In particular, one can consider the following ansatz.
\begin{equation}\label{ansatzI}
\hat{\mathcal{I}}= A_2(t) \hat{\mathcal{O}}_2 + A_3(t) \hat{\mathcal{O}}_3 + A_4(t) \hat{\mathcal{O}}_4 + A_6(t) \hat{\mathcal{O}}_6 + A_7(t) \hat{\mathcal{O}}_7. 
\end{equation}
Putting \ref{ansatzI} in \ref{invariant condition}, one will  have a set of coupled equations of the coefficients. If this set of equation has a solution, then the system will have a LR-invariant of the form \ref{ansatzI}. In our case, the set of equations of the coefficients are the following.
\begin{eqnarray}\label{coefficientequations}
\dot{A}_2=0. \\
\dot{A}_3 + A_2 - \alpha A_6 -\beta A_7 =0. \\
\dot{A}_4 + A_2 - \alpha A_6 -\beta A_7 =0. \\
\dot{A}_6 - \beta A_3 -\beta A_4 =0. \\
\dot{A}_7 - \alpha A_3 -\alpha A_4 =0.
\end{eqnarray}
Here prime represent the derivative with respect to t. The set of equations \ref{coefficientequations} indeed has an unique solution. However for arbitrary time dependence of $\alpha$ and $\beta$ will makes it difficult to obtain the exact solution. For demonstration purpose let us restrict ourselves in constant $\alpha$ and $\beta$. In particular
\begin{eqnarray}\label{parametersA}
A_2 = \alpha_6 \alpha + \alpha_7 \beta . \\
A_3= \frac{1}{2}\alpha_3 + \frac{a_0}{2\alpha\beta}\cosh(2\sqrt{\alpha\beta}t + \theta_0). \\
A_4= -\frac{1}{2}\alpha_3 + \frac{a_0}{2\alpha\beta}\cosh(2\sqrt{\alpha\beta}t + \theta_0). \\
A_6= \alpha_6 + \frac{a_0}{2\alpha\sqrt{\alpha \beta}}\sinh(2\sqrt{\alpha\beta}t + \theta_0). \\
A_7= \alpha_7 + \frac{a_0}{2\beta\sqrt{\alpha \beta}}\sinh(2\sqrt{\alpha\beta}t + \theta_0).
\end{eqnarray}
Where $\left\{\alpha_3, \;\alpha_6,\; \alpha_7,\; a_0,\;\theta_0 \right\}$ are integration constants.\\
Using these values of $A_i$'s in \ref{ansatzI}, one can obtain the explicit form of the invariant operator. The case for time dependent $\alpha$ and $\beta$ can readily be generalized in straightforward manner. However for arbitrary time dependency on  $\alpha$ and $\beta$ will restrict the existence for unitary time dependent canonical transformation (TDQCT), under which we can obtain an equivalent time independent hamiltonian. In the next section we have demonstrated the existence of a class of TDQCT.
\section{Canonical transformation}
First let us recall the action of a time-dependent quantum canonical transformation (QCT) on quantum system \cite{qct1,qct2,qct3,qct4,qct5,qct6,qct7,qct8,qct9}. One can note that the time dependent Schr\"{o}dinger equation reads
\begin{equation}\label{scevolution}
\hat{H}(t) \hat{U}(t) =i\dot{\hat{U}}(t) . 
\end{equation}
 Since under QCT the time remains unaffected in the original system and the transformed system, the settlement of reference-time is very essential for QCT. In the present discussion, we are measuring time from some initial time $t=0$. In particular, evolution operator $\hat{U}(t)$ fulfills the condition $\hat{U}(0) = \hat{\mathbb{I}}$.
Next obvious requirement of  QCT is that the Schr\"{o}dinger equation \ref{scevolution} remains invariant under time-dependent QCT $\hat{\mathcal{U}}(t)$.  These two requirements enforce the following transformation rules.
\begin{eqnarray}
 \hat{U}  :\rightarrow \hat{U}_1 = \mathcal{\hat{U}} (t) \hat{U}(t) \mathcal{\hat{U}}^\dagger(0).\\
\hat{H} :\rightarrow \hat{H}_1 = \hat{\mathcal{U}}\hat{H} \hat{\mathcal{U}}^\dagger -i \hat{\mathcal{U}} \dot{\hat{\mathcal{U}}}^\dagger .
\end{eqnarray}
We are confining ourselves for the PDEM system which possess a close form LR-invariant. Therefore, we are dealing with the following PDEM system
\begin{eqnarray}\label{hforct}
\hat{H}(t)= \hat{p}\hat{\mu}(x,t)\hat{p} + v_{eff}. \\
\mbox{With,}\;\; \mu(x,t)=\frac{1}{2m(x,t)}.
\end{eqnarray}
Where choices of mass $(m(x,t))$ and potentials $(V(x,t))$ are restricted within \ref{choice of m} and \ref{choice of v}. For simplicity let us further restrict ourselves for the case $\beta =0$. That means we are interested for the potential of the form
\begin{equation}
V(x,t)= V_0(t) -\frac{\alpha^2}{c_0(t)+\alpha x} .
\end{equation}
And the form of mass in which we are intersted.
\begin{equation}
m(x,t)= -\frac{1}{2\left(c_0(t)+\alpha x\right)}.
\end{equation}
Now our task is to define a proper time dependent QCT for the Hamiltonian \ref{hforct}. Let us define the following time dependent quantum canonical transformation.
\begin{equation}
\hat{\mathcal{U}} = e^{i\epsilon(t)f(\hat{p})}.
\end{equation}
Where $f(\hat{p})$ is an arbitrary function of $\hat{p}$. One can note the following transformation rules.
\begin{eqnarray}
\hat{p}:\rightarrow \hat{p}_1= \hat{\mathcal{U}} \hat{p}  \hat{\mathcal{U}}^\dagger = \hat{p} . \label{tranformedp} \\
\hat{x}:\rightarrow \hat{x}_1= \hat{\mathcal{U}} \hat{x}  \hat{\mathcal{U}}^\dagger = \hat{x} + \epsilon(t) f'(\hat{p} ) . \label{tranformedx}
\end{eqnarray}
The Hamiltonian is transformed as
\begin{eqnarray}\label{transformedH}
\hat{H}(x,p,t) :\rightarrow \hat{K}(x_1,p_1,t)=  \hat{\mathcal{U}}\hat{H} \hat{\mathcal{U}}^\dagger -i \hat{\mathcal{U}} \dot{\hat{\mathcal{U}}}^\dagger . \nonumber \\
\therefore \hat{K}=-c_0(t)\hat{p}^2 -\alpha \epsilon(t)f'(\hat{p})\hat{p}^2 - \alpha \hat{p}\hat{x}\hat{p} -\dot{\epsilon}(t)f(\hat{p}) + V_0(t).
\end{eqnarray}
We have to choose $\epsilon$ and $f(\hat{p})$ in such a manner so that \ref{transformedH} becomes time independent. As we have already mentioned that we are only interested in the systems which are quadratic in momentum. Therefore, the term $f'(p)p^2$ suggests that
\begin{equation}\label{fp}
f(\hat{p})= \delta_0(t)+\delta_1(t)\hat{p}.
\end{equation}
Using \ref{fp} in \ref{transformedH}, one can note that $\hat{K}$ can be written as
\begin{eqnarray}
\hat{K}=-\left(c_0(t)+\alpha\epsilon(t)\delta_1(t)\right)\left(p+\frac{\dot{\epsilon}(t)\delta_1(t)}{2 \left(c_0(t)+\alpha\epsilon(t)\delta_1(t)\right)} \right)^2 - \nonumber \\
\alpha \hat{p}\hat{x}\hat{p} + \frac{\dot{\epsilon}^2(t)\delta_1^2(t)}{4\left(c_0(t)+\alpha\epsilon(t)\delta_1(t)\right)} -\dot{\epsilon}(t)\delta_0(t) +V_0(t).
\end{eqnarray}
Since, the mass profile and potential functions are given (that means, they can be controlled in experimental set up), so we have to choose the unknown parameters in such a manner that the unknowns can be determined by the experimentally controlable parameters. In our discussion, such controlable time dependent parameters are $c_0(t)$ and $V_0(t)$. Therefore we can choose the following restrictions to make $\hat{K}$ time independent.
\begin{eqnarray}
c_0(t)+\alpha\epsilon(t)\delta_1(t)=-\mu_1 . \label{mu1} \\
\frac{\dot{\epsilon}(t)\delta_1(t)}{2\left(c_0(t)+\alpha\epsilon(t)\delta_1(t)\right)} = \mu_2 . \label{mu2} \\
V_0(t)-\dot{\epsilon}(t)\delta_0(t)-\mu_1\mu_2^2 =\mu_3 . \label{mu3}
\end{eqnarray}
Where $\mu_1 \neq 0,\;\mu_2 \neq 0 \; \mbox{and}\; \mu_3$ are constants and $\hat{K}$ becomes
\begin{eqnarray}
\hat{K}=\mu_1 (p+\mu_2)^2-\alpha \hat{p}\hat{x}\hat{p} + \mu_3.
\end{eqnarray}

Each value of $\mu_i$'s will correspond to a QCT.
Solving Eq.\ref{mu1}, \ref{mu2} and \ref{mu3}, one can write down the explicit form of $\epsilon(t),
\;\delta_0(t), \; \delta_1(t) $. 
\begin{eqnarray}
\epsilon(t)=\epsilon_0 \exp\left(2\alpha\mu_1\mu_2 \int^t \frac{d\tau}{\mu_1+ c_0(\tau)}\right). \label{epsilonform} \\
\delta_1(t)= -\frac{c_0(t)+\mu_1}{\alpha\epsilon(t)}. \label{delta1form} \\
\delta_0(t)= -\frac{\left(V_0(t)-\mu_3-\mu_1\mu_2^2\right)}{2\mu_1\mu_2}\delta_1(t). \label{delta2form}
\end{eqnarray}
Now one can utilize the constructed time-independent hamiltonain $\hat{K}$ to construct the wave functions in transformed space by methods like SUSY formalism \cite{susy1,susy2,susy3,susy4,susy5,susy6}. By inverse transformation the wave functions in original space can be easily obtained. 
\section{Discussion}
We have shown that a time-dependent quantum canonical transformation (TDQCT) can be constructed for the system with position-dependent effective mass (PDEM). However, the class of allowed PDEM for which the close form Lewis-Riesenfeld -Ermakov's invariant operator (LRIO) can be constructed is limited. It seems that the existence of a natural inertia potential corresponding to a PDEM restricts the class of external potentials for which LRIO will be of the close form.
Since the time dependent potential function $V(t)$ is arbitrary and the inclusion of $c_0(t)$  in both mass profile and external potential is also arbitrary so our approach is fairly general. Hence it can be utilized to model the physically important systems like semiconductors, especially to study its electronic properties. \\
Most striking point of our findings is that, the allowed interaction and mass profile are mathematically similar to each other for close form LRIO. Moreover, they looks like Coulomb's potential with time dependent shifted origin. Since the inertia measure will be different for different spatial points for a particle with PDEM, spacial points do not appear in identical footing to a particle with PDEM. Therefore the appearance of $V_c(x,t)\sim \frac{2\alpha^2 m_0}{1+l^{-1}x}$ in allowed potential suggests that there must be some relationship between the electric charge and spacial points. One may be tempted to claim that the findings of the present article validate the fact of considering the electric charge as another component of the particle momentum in an additional dimensional space–time. Thus our findings is a supporting evidence for a possible unification of Newton's and Coulomb's forces \cite{coulombnewton,klein,klein1}.
For simplicity we have restricted ourselves for the constant structure constants $\alpha$ and $\beta$. However, the case of time dependent $\alpha$ and $\beta$ can easily be generalized. To avoid the fractional power of the operators  in TDQCT we have restricted ourselves for the case $\beta=0$.

\end{document}